# Nature and Quantization of the Proton Mass: An Electromagnetic Model

G. Sardin

Applied Physics Department, University of Barcelona

Abstract

A method for quantization of the proton mass is here addressed, which provides a plausible explanation for the origin of mass and leads to the unification of mass and electric charge through their coupling. By means of an electromagnetic approach, the calculated mass of the proton closely approximates its experimental value and does so with dependence on a single parameter. That is to say, the proposed fundamental system provides a way to comprehend the source of mass as a property of the structure of elementary particles. It brings a new tool to the task of gaining insight into the proton mass and to unravelling the enigma of proton stability. The inner energy of elementary particles, or equivalently their mass, is surmised here to have electrodynamic roots, deriving from the dynamics of a single or pair of electric charge(s) shaping out their structure. Mass appears as the quantized balance of two inner energies which conform collapsing action and retentive reaction. Charge and mass are not taken as independent entities as in the traditional mode, instead mass appears as a by-product of the charge structural dynamics, as does the magnetic moment. The proposed model clearly requires a degree of willingness to consider possibilities not accounted for within the framework of the Standard Model. So, this proposal is addressed to those who are open to inspect a different look at the structure of elementary particles and disposed to compare the two approaches, standing out of doctrinal captivity.

Introduction

In an earlier paper (1), the fundamentals of the proposed approach to the structure of elementary particles have been reported, and a few developments along the same lines are cited in references (2-5). The generic structure of all elementary particles is regarded as being defined by an orbital ruled by a structural wavefunction $\Psi$, which determines their quantum state $|\Psi>$ and all their properties: mass, magnetic moment, mean size, specific structure, and so on. This viewpoint is here applied to the proton, the only stable charged particle along with the electron, which combine in the form of the hydrogen atom to make up the content of all stars and thus the major constituent of the visible universe. Unstable particles are viewed as off equilibrium structural quantum states whose life-time is extremely short (typically from $10^{-7}$ to $10^{-23}$ s), and whose roles are subsidiary even though there is a great variety.

At the beginning of 1900, first attempts at physical theory seeking to elucidate the origin of mass and inertia formed the basis of what came to be known as classical electromagnetic mass theory. At that time the theory was exclusively applied to the electron and was thus referred to as classical electron theory. Some inner difficulties and the advent of the theories of relativity and quantum mechanics led to its shelving. However, despite the passage of time, the nature of mass and inertia remained unanswered with some credibility, i.e. without having to align doctrinally with all the weirdness of the standard model, and with its hypothetical and surrealistic Higgs boson and Higgs field on which the origin of mass would depend, or to rely on the still odder super-symmetric version of that model, with multiple Higgs bosons and fields.

In view of the poor reliability of such erratic approaches let's recover a long overdue touch of realism and rely back on a physical body, the electric charge, which offers the irresistible appeal of being an observable, in contrast with quarks, gluons, Higgs bosons and Higgs fields, none of them having ever been observed, and whose presumed existence only holds on the feeding of conjectural elementary particles and cosmological models, made out of an abusively speculative mathematical texture.

Apart from its credibility, the value itself of the Higgs approach is doubtful since unable to provide unification of mass and electric charge, leaving so these two crucial physical entities as cut off items. In view of the poor reliance on mass rooted on the airy Higgs field, and its inaptitude to link electromagnetism and gravity, let's regain some good sense and retake more grounded foundations sitting on the integer electric charge, which at least is observable. So, let's rescue the electromagnetic theory of mass from its oblivion and give back some credit to those physicists who inferred to mass an electromagnetic origin.

This assumption has a long history, to which are associated such names as Abraham, Bakker, Fermi, Feynman, Harvey, Heaviside, Hughes, Kaufman, Kwal, Langevin, Lorentz, Mandel, Poincaré, Petkov, Pryce, Righi, Rohrlich, Searle, Van der Togt, and Wilson, to mention only a few. Already in 1881, J. Thompson was attempting to explain matter as an electromagnetic phenomena. In 1897, G. Searle depicted inertia in purely electromagnetic terms. In 1905, P. Langevin (6) inspected the possibility of mass having an electromagnetic origin. R.P. Feynman (7), in the chapter entitled "*Electromagnetic Mass*" of his book "*Lectures on physics, Volume II*", says in § 28, p.2: "*It might, in fact, be that the mass is just the effect of electrodynamics*". In the book "*From Paradox to Paradigm*", J. Bakker (8) *also* addresses the electromagnetic nature of mass. C. van der Togt (9) assumes the "*Equivalence of Magnetic and Kinetic Energy*". R. Stevenson and R.B. Moore (10), in the book "*Theory of Physics*", assert that there is experimental evidence of electromagnetic mass. Unfortunately, the rise of the theory of general relativity put an end to this incompletely inspected conceptual framework. The historical development of the classical electromagnetic mass theory is addressed in references (12-15).

However, the electromagnetic mass conjecture mostly refers to the nature of inertial mass, scarcely ever to the rest mass. The concept has not been applied to elementary particles as a unified phenomenology nor has it been developed as an electromagnetic model of their structure and associated mass, as it is put in reference (1). Here, the nature of the rest mass of elementary particles is brought into play and is specifically applied to the proton. Our aim, then, is to corroborate the diverse premises concerning the electromagnetic origin of mass and to extend them to the whole mass, inertial and intrinsic. However, we will not apply here the electromagnetic theory of mass to the electron, as done more or less shrewdly by precursors, but to the derivation of the proton mass by means of an electromagnetic model of its structure. From it, mass appears as an outcome of the electric charge confined dynamics, which shapes the structure of all elementary particles (1,7).

Nature of the proton mass

In phenomenological terms, the proton structure is considered to be embodied by the orbital of a unitary electric charge, ruled by a wavefunction Ψ fixing the proton structural quantum state $|\Psi\rangle$ and all derived properties. In that generic framework, the proton mass is understood to arise from the equilibrium between two antagonist energies – one compressive and the other one expansive – which define its structural net energy, that is to say, its rest mass. Even though a formal treatment would certainly appeal to QEM, a quanto-mechanical treatment is nevertheless not imperative to support quantization of the proton mass. A semi-classical formulation allows it in a first approximation, which presents the advantage of offering a concrete electromechanical understanding of the proton structure, in like fashion as does the Bohr model for the atomic structure.

The pursued conceptual approach is based on an elemental gyrator system consisting in an integer electric charge self-confined within a closed path that defines the structural orbital. The self-rotation of the electric charge generates two antagonist forces whose relationship determines the equilibrium state of this elemental gyrator system. In such a scheme the self-trapping at the Fermi scale of the charge is due to the interaction with its own magnetic field that subjects it to a centripetal force $F_\downarrow = (e^2/r^2)(v^2/c^2)$, which in turn provides feedback into the rotation and thus to the self-confinement. The so-called self-interaction is a basic precept at the Fermi scale. On the other hand, it will be assumed that the rotation of the electric charge simultaneously creates an antagonist centrifugal force in order to avoid the system collapse, expressed as: $F_\uparrow = (mv^2)/r$. This would lead to a type of fundamental action and reaction at the most elemental level. It rises that the assumption of an emerging resistance to the system collapse, through the reactive centrifugal force, implies the creation of the magnitude called mass as a by-product of the charge dynamics. An essential precept of the theory of the electromagnetic nature of mass relies on the interaction of the electric charge with its own magnetic field (7,10). Let's now specify the formulation of the proton mass.

## Quantization of the proton mass

Let us assume that an integer electric charge e rotating along an orbit of radius r generates two antagonist forces, one centripetal (action) and the other one centrifugal (reaction), that fix its equilibrium state, and let's start with the non relativist formulation. In the unit system cgs, the expression of the centripetal force is:

$$F_\downarrow = e(v/c)H = e(v/c)[(e/r^2)(v/c)] = (e^2/r^2)(v^2/c^2)$$

where $H = (e/r^2)(v/c)$, and is here the own magnetic field of the gyrator system. That is to say, the rotation of the charge generates a magnetic field, which interacts with the charge and so confines it through the emergence of a Lorentz centripetal force. This is a non-classical effect that applies locally at the Fermi scale, and is mostly a quantum feature. It generically belongs to the so-called "self-interaction". $F_\downarrow$ can be suitably rewritten in the following form:

$$F_\downarrow = [(e^2/m_0c^2)/r^2](v^2/c^2)\,m_0c^2 = (r_0/r^2)(v^2/c^2)\,m_0c^2$$

0n its turn the non relativist expression of the centrifugal force is:

$$F_\uparrow = (m_0v^2)/r = [(m_0c^2)/r](v^2/c^2) = (1/r)(v^2/c^2)\,m_0c^2$$

So, their difference $\Delta F$ is equal to:

$$\Delta F = F_\downarrow - F_\uparrow = (r_0/r^2)(v^2/c^2)\, m_0 c^2 - (1/r)(v^2/c^2)\, m_0 c^2 = (r_0/r^2 - 1/r)(v^2/c^2)\, m_0 c^2$$

where $r_0 = e^2/m_0 c^2$ and corresponds to the so-called classical electron radius. Let's highlight that for $r = r_0$ the system is at equilibrium since $\Delta F = 0$, and thus its energy is null. However, if the radius r drifts from its equilibrium value ($r \neq r_0$), the energy involved in the variation of r from $r_0$ to $r_1$ is:

$$E = \int \Delta F\, dr = \int [(r_0/r^2 - 1/r)(v^2/c^2)]_{r_1}^{r_0}\, m_0 c^2\, dr = [(r_0/r - \log(r))(v^2/c^2)]_{r_1}^{r_0}\, m_0 c^2$$

$$E = [(r_0/r_1 - 1) - \log(r_0/r_1)](v^2/c^2)\, m_0 c^2$$

The corresponding relativistic formulation is:

$$E = [(r_0/r - 1) - \log(r_0/r)](\gamma - 1)\, m_0 c^2$$

in which the previous non relativistic term ($v^2/c^2$) has been replaced by ($\gamma-1$) instead of just $\gamma$ since for $v = 0$ the energy E must be null.

Let's now state a quantitative relationship between the speed v of the electric charge acting as the structural carrier of elementary particles and the mean radius r of its orbital. We assume a coupling between v and r such as: $(\gamma-1) = (\alpha\, r_0/r)^{-2}$, i.e.:

$$\gamma = 1 + (\alpha\, r_0/r)^{-2}$$

where $\alpha^{-1} = \hbar c/e^2 = 137.036$ (inverse fine-structure constant). It comes out that for $r \to 0$, $v \to 0$ and for $r > 0.8$ Fm, $v \to c$ asymptotically. When the system is at equilibrium, i.e. for $r = r_0$, v is extremely close to c, with 8 identical digits (Fig.1).

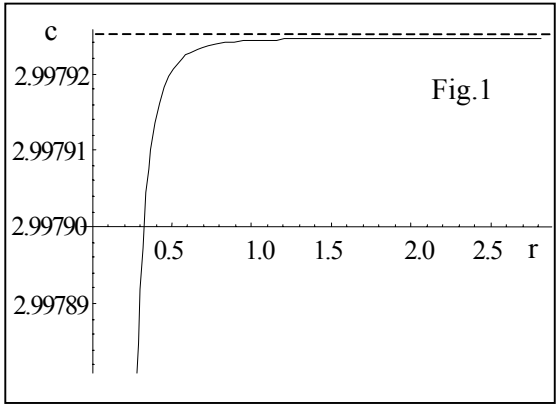 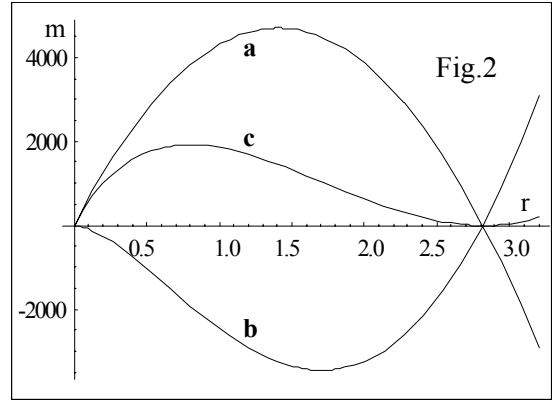

Fig.1: Coupling between v and r. Note that v is almost constant for r > 0.8 Fm and almost equal to c without ever reaching it.

Fig.2: Evolution of the energy E of the elemental electrodynamic system vs. its radius r. The curves represent: (a) the centripetal energy, (b) the centrifugal energy and (c) the resultant net energy, which on its turn expresses the mass of the particle. The curve of the net structural energy provides a single maximum value, identified as corresponding to the proton mass in view of its proximity to its experimental value and also due to the fact that it is the only stable highly massive charged particle.

So, for v and r being coupled, the corresponding expression of the energy E is:

$$E = [(r_0/r - 1) - \log(r_0/r)][\alpha (r_0/r)]^{-2} m_0 c^2$$

It comes out from this expression of E that the sole determination of the radius fixes the energy of the system (Fig.2).

As an attempting illustration (Fig.3), a few unstable particles have been added on the section of the curve for which r > 2.8 Fm. In that part of the curve the offset between constricting and expanding energies increases rapidly along with the radius and thus their structure is increasingly off equilibrium. Inversely, through transition to a smaller radius they decay to a lower energy state. Unstable particles may however not necessarily fit on the curve since they do only if their structural energy has no oscillatory component. For example, the muon lies off the curve because its energy is vibrational, i.e. due to the oscillation of its radius and not to a change of radius with respect to that of the electron. So, the muon can be seen as a vibrating electron. For its part, the electron is stable because it is at the curve minimum, i.e. it represents the ground state. The proton, which appears as the only particle with r < 2.8 Fm, is stable despite being at the top of the left part of the curve where the offset between the two antagonist structural energies, compressive and expansive, is maximum. This is due to the fact that mass and magnetic moment are coupled, implying the proton to be trapped at the bottom of a deep magnetic well, from which to come off would need more inner energy than it has, and thus it cannot decay (Fig.4).

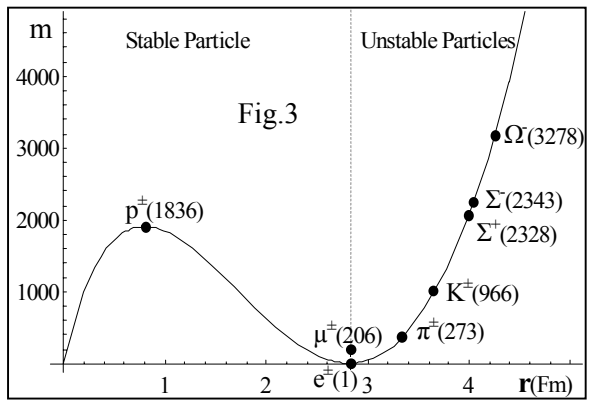
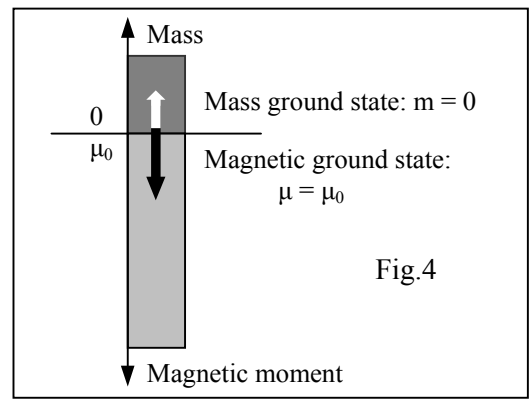

Fig.3: Structural energy level of a few elementary particles.

Fig.4: Since mass and magnetic moment are inversely coupled, if m increases it does at the expense of µ that must decrease, and vice versa.

## Quantization of the energy:

(a) *Method of the maximum*

In Fig.2, let's note that the most relevant feature of the evolution of E upon the variation of r from its initial value $r_0$ to 0, is that it exhibits a maximum value of 1912 $m_0c^2$, close to the proton mass (1836 times that of the electron) for a radius of about 0.80 Fm, a value that in turn is close to the average proton radius of ~1 Fm, and of 0.85 Fm from scattering experiments (16). So, the quantization of the electrodynamic system proposed may be considered to be taken as an approximation to the proton mass and size, opening the door to an understanding of its structure and to part of its properties, since expressing: (a) the electromagnetic nature of the proton mass, (b) the quantization of its mass, (c) the cause of its stability, in view that, from the proposed electrodynamic model, the proton appears trapped in an electromagnetic well.

The criteria used to determine mass has been to select the maximum value of the net energy E vs. the radius r. However, it has been seen that the corresponding value does not perfectly match with the expected one. The reason for it may eventually turn out to be that each one of the three energies reaches maximum value for a different value of r. It could be extrapolated from this fact that the system may actually reflect compliance to some kind of compromise between the value of r from the collapsing, expansive, and net energy, respectively equal to 1.41, 1.71 and 0.80 Fm.

The experimental value of the proton radius turns around 1.0 Fm, with some degree of uncertainty due to the experimental difficulties inherent in inspecting its structure. Just to set limits, let's mention that the size of nucleons deduced from atomic nuclei is about 1.4 Fm, a value constituting a higher limit. It is also known that at a distance of 0.5 Fm the nuclear forces become repulsive, so this value sets a lower limit to nucleon size. High-energy scattering experiments have provided a value of 0.85 Fm with a spread of 0.15 Fm (16). So, proton size appears somewhat dependant on the type of experiment from which it has been inferred, but there is a wide consensus on a referential mean value of ~1 Fm.

(b) *Method of the Magnetic Moment*

In view that the method of the maximum value of the energies vs. the radius leaves some ambiguity about the most adequate value of the radius to be picked up, and that the model does not provide concrete hints about it, let's thus appeal to another method based on the magnetic moment of the proton, which offers the advantage of this magnitude being measured with an extreme accuracy. So, to improve the calculated value of the proton mass let's appeal to its magnetic moment. Its standard quantum expression is:

$\mu = g\,(e\hbar/2m_p c)$

where $g = \mu_p/\mu_N = 2.792847386$ and expresses the ratio between the proton effective magnetic moment and the nuclear magneton. As well known, the proton magnetic moment is g times higher than the expected value calculated from its mass, without the ad hoc correction corresponding to the g factor. We have advanced elsewhere that, in order to obtain the correct value of the magnetic moment, the specific size of the particle considered should be accounted for, and the physical meaning of the corrective g factor would correspond to the ratio between the electron classical radius and the effective proton radius, i.e. $g = r_0/r_p$. This relation has been advanced in reference (2).

Knowing that $r_0 = 2.817941$ Fm, hence $r_p = r_0/g = 2.817941 / 2.792847$, and thus:

$r_p = 1.008985$ Fm

This very precise value for the mean electromagnetic radius of the proton will then be used to calculate the proton mass from the expression E of the inner proton energy:

$$E = [(r_0/r_p - 1) - \log(r_0/r_p)][\alpha\,(r_0/r_p)]^{-2}\,m_0 c^2 = 1843.66\,m_0 c^2$$

value that is to be compared with the experimental one for the proton mass of 1836.15 $m_0 c^2$. The relative spread between the calculated and experimental value is of only 0.4 %.

It should also be pointed out that the derived expression E of the inner electrodynamic energy applies also to the electron, the only stable charged particle along with the proton. In effect, the curve of structural energy exhibits a minimum (Fig.1) at $r = r_0$, at which point the net energy E is null. This may explain why the electron is often considered to be structure-less, since its dynamical structure is energy-less and thus undetected (from high-energy scattering). However, this point-like feature makes sense with respect to the co-existent classical electron radius. The electron punctual and bulky aspects, which heretofore have been considered to be mutually exclusive, are no longer so. As well, the fact that the net dynamical energy of the electron structure is null implies that its mass is not electrodynamic in origin, but electrostatic instead. In counterpart, its magnetic moment would rely on its structural electrodynamics and thus on its classical radius, so both features acquire full coherence.

Let us note also that, in applying the same type of quantization as that done for the cgs formulation of the energy E to the classical expression of the magnetic moment, we obtain its exact quantum expression.

$$\mu = 1/2\,e\,r\,(v/c)$$

In the classical formulation of E we have replaced the term $(v^2/c^2)$ by $(\alpha\,r_0/r)^{-2}$, so let's now, similarly supplant in $\mu$ the term $(v/c)$ by $(\alpha\,r_0/r)^{-1}$. Thus:

$$\mu_0 = 1/2\,e\,r_0\,(\alpha\,r_0/r)^{-1} = 1/2\,\alpha^{-1}\,e\,r_0\,(r/r_0) = 1/2\,e\,(\hbar c/e^2)(e^2/m_0 c^2) = 1/2\,(e\hbar/m_0 c)$$

since for the electron $r = r_0$. So, the application of the quantum factor $(\alpha\,r_0/r)^{-1}$ to the classical formulation of $\mu$ gives the right quantum expression for the magnetic moment of the electron. For the proton it gives:

$$\mu_p = g\,(m_0/m_p)\,\mu_0 = (r_0/r_p)(m_0/m_p)\,\mu_0 = 1/2\,(r_0/r_p)(m_0/m_p)(e\hbar/m_0 c)$$

which is the expression for the proton magnetic moment, in which g is interpreted as expressing the ratio between the electromagnetic radius of the electron and the proton. Hence, this expression of $\mu_p$ makes it evident that the proton and the electron are linked through the same structure but are in different states, and also that the proton magnetic moment is proportional not only to the mass ratio $m_0/m_p$ but also to the size ratio $r_0/r_p$, which appears quite logical.

The reason why the expression $\mu = e\hbar/2mc$ works out for the electron but not for the proton, is because it presupposes that the particle classical radius is in both cases equal to the classical electron radius, i.e. $r_0 = e^2/m_0 c^2$. In effect, the quantum expression $\mu = 1/2\,(e\hbar/mc)$ can be directly derived from its classical homologue $\mu = 1/2\,e\,r\,(v/c)$, by multiplying it by the inverse fine-structure constant $\alpha^{-1}$ and equating v to c:

$$\mu(\text{quantum}) = \mu(\text{classical})(\text{inverse hyperfine structure constant}) = 1/2\,e\,r\,(v/c)\,\alpha^{-1}$$

Applied to the electron it gives:

$$\mu_0 = 1/2 \, e \, r_0 \, \alpha^{-1} = 1/2 \, e \, (e^2/m_0c^2)(\hbar c/e^2) = e\hbar/2m_0c$$

which is the standard expression of the electron magnetic moment. If applied to the proton, it fails because the radius $r_0$ is retained and so it inadvertently infers that both particles have the same classical radius $r_0$, which is not the case.

In a transition from $r_0$ to $r_1$ the system absorbs a magnetic moment $\Delta\mu = - (\mu_0 - \mu_p)$. This drop $\Delta\mu$ from the initial value $\mu_0$ represents the magnetic well in which the proton is trapped, and could provide the explanation for its stability. So, the proton has a deficit of magnetic moment $\Delta\mu = - 9.25919 \, 10^{-24}$ J.T$^{-1}$ with respect to the system ground state represented by the electron, i.e. a deficit of 99.85 % of $\mu_0$.

In effect, in the proton state $|\Psi_p\rangle$, the system has a mass of 1836.15 $m_0$ relative to its ground state $|\Psi_0\rangle$, but it has a magnetic moment deficit:

$$|\Delta\mu| = \mu_0 - \mu_p = 9.27329 \, 10^{-24} - 1.41049 \, 10^{-26} = 9.25919 \, 10^{-24} \text{ J.T}^{-1}$$

The point is that this magnetic moment deficit can be transcribed into mass equivalence. The formula used for the $\mu$ to m conversion is:

$$\mathbf{E = 2 \, \mu_0^{\,2} \, r_0^{\,-3}}$$

which expresses that at equilibrium (i.e. for $r = r_0$) the energy equivalence due to the system inner electro-dynamics is equal to the density of the squared magnetic moment, and its value is:

$$E = 2 \, (e\hbar/2m_0c)^2 \, (e^2/m_0c^2)^{-3} = 1/2 \, (\hbar c/e^2)^2 \, m_0c^2 = 9389.43 \, m_0c^2$$

Therefore, in view of its magnetic moment deficit, the proton would need to have a mass of 9389.43 $(\Delta\mu/\mu_0) \, m_0 = 9375.15 \, m_0$ to be able to get out of the magnetic well, but its mass is only of 1836.15 $m_0$, so it falls short of 7539.0 $m_0$. This value represents the mass to be transformed into magnetic moment for the proton to be able to get out of the well. Let's remind that $\mu$ and m are inversely coupled, so if m increases, it does at the expense of $\mu$ that necessarily decreases, and vice versa (4).

Discussion and conclusion

The aptitude of this approach in providing a new look into the structure of the proton, giving a novel access to its mass and more generally opening a new horizon to grasp the nature of elementary particles, is to be pondered. In effect, an electromagnetic approach to the origin of mass has allowed us to calculate the proton mass. It derives from the drift of the radius r from its classical equilibrium value $r_0$ to a quantum equilibrium value $r_p$, at which point the net structural energy E reaches a maximum and offers a relatively adequate approximation to its experimental mass. We emphasize here that such a rudimentary model of the proton structure, which only uses a semi-classical framework without introducing arbitrary touches, provides conceptual grounds and clues for a more sophisticated approach based on QED, surely more genuine and reliable than QCD. As a standing out corollary, let's be aware of the fact that, since the mass of particles relies on their structural energy, the breakdown of their structure implies the evanescence of their mass and consequently of the associated gravitational field.

It has been shown that the model provides the mass of the proton with a better accuracy when the radius is defined from the magnetic moment of the proton. Let's highlight the conceptual coherence of the interpretation of the physical meaning of the g factor as corresponding to the ratio $r_0/r_p$. The fact that the use of the corresponding value of the electromagnetic radius $r_p$ provides the proton mass, strengthens the grounds of the model used for the structure of elementary particles and in particular that of the proton. We should note, however, that the model is limited by virtue of its being only two-dimensional while the proton, even though not having an exact spherical symmetry, is nevertheless three-dimensional. The model clearly needs to be enhanced but it already opens doors to a new conception of elementary particles. An improved approach might likely be purely quanto-mechanical, however let's remind that the Bohr model for the H atom gave surprising good results although being also two-dimensional, despite the H atom being three-dimensional. It appears thus that the rest energy can be evaluated without the requirement of knowing the exact distribution of the orbital embodying the structure.

Let us stress that the way to calculate the proton radius $r_p$ – on one hand from the quantum formulation of its magnetic moment $\mu_p$ and the assumption that the factor $g = \mu_p/\mu_N = 2.79285$ expresses the ratio $r_0/r_p$, and on the other hand from its experimental mass when using the expression of E of the structural model proposed – provides respectively the value $r_p = 1.01$ Fm and 1.02 Fm, i.e., a relative spread of 1%. Let's highlight that through the only knowledge of the proton radius the model provides at once its magnetic moment and its mass.

Let's also point out that it uses only one variable, the radius, and not a single parameter of adjust. Compare this with the 20 parameters used by the Standard model. Defenders, in a forty years long systematic forward escape, and in a vain effort to offset its decreasing credibility along with increasing unmanageability, have raised it to such a dogma that followers have ended up losing the sagacity to ponder its very basis, starting with the much unfortunate partition of the electric charge. Our strategy consists in the search of simplicity, while the Standard Model clearly expresses the cult to complexity and the complaisance in further increasing it in front of any emerging difficulty, as proves the advent of the Super Symmetric Standard Model. With awareness, just freely choose between these two intellectual standpoints. We welcome contributions to further improvements of the rudimentary but nevertheless operative model outlined.

## Acknowledgements

All my thanks to Don Johnson for his kind revision and styling of the English writing.

## References

(1) G. Sardin, *Fundamentals of the Orbital Conception of Elementary Particles and of their Application to the Neutron and Nuclear Structure*, Physics Essays, 12, 2 (1999)
http://uk.arxiv.org/ftp/hep-ph/papers/0102/0102268.pdf
(2) G. Sardin, *Unification of the electron and proton* http://www.terra.es/personal/gsardin/news17.htm

(3) G. Sardin, *Nature of the muon mass,* http://www.terra.es/personal/gsardin/news16.htm
(4) G. Sardin, *A new universal constant Q,* http://www.terra.es/personal/gsardin/Q.doc
(5) G. Sardin, *The hydrogen atom Q charge,* http://www.terra.es/personal/gsardin/news19.htm
(6) (a) P. Langevin, *La physique des électrons*, Rapport au Congrès international des Sciences et Arts, Saint Louis (1904). Revue générale des sciences pures et appliquées 16, 257–276 (1905)
(b) P. Langevin, *L'inertie de l'énergie et ses conséquences*, J. de Physique 3, 553-582 (1913)
(c) P. Langevin and H. Abraham, *Mémoires relatifs à la physique, ions, électrons, corpuscules*, Gauthier-Villars, Paris (1905)
(7) R.P. Feymann, *Lectures on Physics, Vol.II (The electromagnetic mass),* Addison-Wesley, N.Y.(1964)
(8) J. Bakker, *From Paradox to Paradigm*" (1999)
http://www.paradox-paradigm.nl/The%20equivalence%20of%20magnetic%20and%20kinetic%20energy.htm
(9) C. van der Togt, *Equivalence of Magnetic and Kinetic Energy*
http://www.paradox-paradigm.nl/Van_der_Togt_equiv2ckw.pdf
(10) R. Stevenson and R.B. Moore, *Theory of Physics*, W.B. Saunders, London (1967)
(11) V. Petkov, *Origin of Inertia,* http://alcor.concordia.ca/~vpetkov/inertiagrav.html
(12) F. Rohrlich, Am. J. Phys. 28, 63 (1960); Classical Charged Particles, Addison-Wesley, N.Y. (1990)
(13) J. W. Butler, Am. J. Phys. 37, 1258 (1969)
(14) A. Pais, *The Early History of the Theory of the Electron: 1897–1947*, edited by Abdus Salam and Eugene P. Wigner, Cambridge: Cambridge University Press (1972)
(15) P. Pearle, *Classical Electron Models*, in *Electromagnetism:Paths to Research*, edited by Doris Teplitz, Plenum Press, New York (1982)
(16) Sangita Haque, L. Begum, Md. Masud Rana, S. Nasmin Rahman and Md. A. Rahman, *Determination of proton size from $\pi^+ p$ and $\pi^- p$ scattering at $T(\pi^\pm) = 277-640$ MeV*
http://www.ictp.trieste.it/~pub_off/preprints-sources/2003/IC2003052P.pdf

ABSTRACT: The pion-nucleon interaction above the Δ(1232) resonance and in the region of low-lying pion-nucleon resonances is studied. π± p elastic scattering at T(π±) = 277-640 MeV characterized by diffraction maxima and minima has been analyzed through the strong absorption model due to Frahn and Venter. ***The proton radius is determined from the best fit values of the cut-off angular momentum to be 0.85 fm with a spread of 0.15 fm. The higher energy pions scan a lower value while the lower energy pions yield a higher value for the size of the proton. The energy averaged radius of the proton size of 0.85 fm obtained in the present analysis is in excellent agreement with proton charge radius of 0.86 fm quoted in the literature.***